\documentclass[10pt,journal,compsoc]{IEEEtran}
\ifCLASSOPTIONcompsoc
  \usepackage[nocompress]{cite}
\else
  \usepackage{cite}
\fi
\ifCLASSINFOpdf
\else
\fi
\usepackage{array}
\usepackage{amsmath}
\usepackage{amsbsy}
\usepackage{amssymb}
\usepackage{algorithm}
\usepackage{algpseudocode}
\usepackage{float}
\usepackage{listings}
\usepackage{url}
\usepackage{hyperref}
\usepackage{graphicx}
\usepackage{todonotes}
\usepackage{enumitem}
\usepackage{tikz}
\usepackage{subcaption}

\usepackage{tikz-cd}

\graphicspath{{./img/}}

\begin{document}
%

\title{Modeling Penetration Testing with Reinforcement Learning Using Capture-the-Flag Challenges: Trade-offs between Model-free Learning and A Priori Knowledge}

%
%
%
%

\author{Fabio~Massimo~Zennaro~
        and~L{\'a}szl{\'o}~Erd{\H o}di
\IEEEcompsocitemizethanks{\IEEEcompsocthanksitem F.M. Zennaro and L. Erd{\H o}di are with the Department of Informatics, University of Oslo, 0316 Oslo, Norway\protect\\
E-mail: fabiomz@ifi.uio.no, laszloe@ifi.uio.no}}

%
%

\markboth{}
{Shell \MakeLowercase{\textit{et al.}}: Bare Demo of IEEEtran.cls for Computer Society Journals}
%



\IEEEtitleabstractindextext{%
\begin{abstract}
Penetration testing is a security exercise aimed at assessing the security of a system by simulating attacks against it. So far, penetration testing has been carried out mainly by trained human attackers and its success critically depended on the available expertise. Automating this practice constitutes a non-trivial problem, as the range of actions that a human expert may attempts against a system and the range of knowledge she relies on to take her decisions are hard to capture. In this paper, we focus our attention on simplified penetration testing problems expressed in the form of capture the flag hacking challenges, and we analyze how model-free reinforcement learning algorithms may help to solve them. In modeling these capture the flag competitions as reinforcement learning problems we highlight that a specific challenge that characterize penetration testing is the problem of discovering the structure of the problem at hand. We then show how this challenge may be eased by relying on different forms of prior knowledge that may be provided to the agent. In this way we demonstrate how the feasibility of tackling penetration testing using reinforcement learning may rest on a careful trade-off between model-free and model-based algorithms. By using techniques to inject a priori knowledge, we show it is possible to better direct the agent and restrict the space of its exploration problem, thus achieving solutions more efficiently. 
\end{abstract}

}

\maketitle

\IEEEdisplaynontitleabstractindextext

%
\IEEEpeerreviewmaketitle

\IEEEraisesectionheading{\section{Introduction}\label{sec:introduction}}

%
%
%
%
\IEEEPARstart{S}{ecuring} modern systems and infrastructures is a central challenge in computer security. As an increasing amount of data and services are delivered through electronic platforms, guaranteeing their correct functioning is crucial for the working of modern society. 

A traditional approach to evaluating security adopts a defensive stance, in which systems are analyzed and hardened from the point of view of a defender. An alternative pro-active perspective is offered by an offensive stance. \emph{Penetration testing} (PT), or \emph{ethical hacking}, consists in performing authorized simulated cyber-attacks against a computer system, with the aim of identifying weaknesses and assessing the overall security. The usefulness of offensive security as a tool to discover vulnerabilities is undisputed \cite{stefinko2016manual}.
PT, though, is a complex and costly activity, requiring relevant knowledge of the target system and of the potential attacks that may be carried against it. Thus, in order to produce relevant insights, PT needs experts able to carefully probe a system and uncover known and, ideally, still unknown vulnerabilities.

A way to train human experts and allow them to acquire ethical hacking knowledge is offered by \emph{capture the flag} competitions (CTF). In a CTF, participants are given the opportunity to conduct different types of real-world attacks against dedicated systems, with the aim of exploiting vulnerabilities behind which they can collect a flag. A CTF is a simplified and well-defined model of PT, usually designed as an educational exercise.

Software applications have been developed to automate some aspects of PT, but they mostly reduce to tools that carry out specific tasks under the direction of a human user. Traditional approaches from artificial intelligence, such as \emph{planning}, were also deployed in the hope of further automating PT through the generation of attack plans \cite{hoffmann2015simulated}; however, human input is still critical to model the context and the target system, and to finally derive conclusions about the actual vulnerabilities. 

Recent advances in artificial intelligence and machine learning may offer a way to overcome some of the current limitations in automating PT. In particular, the paradigm of \emph{reinforcement learning} (RL) \cite{sutton2018reinforcement} was proven to be a versatile and effective method for solving complex problems involving agents trying to behave optimally in a given environment. RL applications embrace a large number of algorithms and methods with varying degrees of computational and sample complexity; in particular \emph{model-free algorithms} rely on minimal a priori knowledge of an environment, and they allow to train an agent simply by interacting with the environment, analogously to the way a player may interact with a game to discover its solution. Indeed games have provided for a long time an excellent benchmark for RL, and model-free methods have achieved state-of-the-art performances in solving many complex games, ranging from traditional Go \cite{silver2017mastering} to modern Atari games \cite{mnih2015human,badia2020agent57}.

These developments suggest the possibility of adopting model-free RL for tackling the PT problem. As a form of gamification of PT, CTFs provide an ideal setting for deploying RL algorithms and training agents that, in the long run, may learn to carry out complete PT independently of human supervision. This idea is not new, and it was in fact spearheaded some years ago by DARPA, which hosted in 2016 the Cyber Grand Challenge Event, a cyber-hacking tournament open to artificial agents trained using machine learning \cite{CyberGrand}.

In this paper, we address the question of the extent to which model-free RL algorithms may be used to solve CTF challenges.  While adopting model-free RL to solve CTF problems may seem a perfect fit, we highlight the specific criticality inherent in PT: obscurity, that is the difficulty in discovering the structure underlying a CTF problem. This may be due either to a CTF system that prevents any sort of information leak or to the presence of defence mechanisms that adapt the target system according to the actions of the agent. We analyze these problems experimentally, considering \emph{high-entropy} and \emph{dynamic} scenarios and evaluating how different RL techniques, such as \emph{lazy loading}, \emph{state aggregation}, and \emph{imitation learning}, may help in tackling these challenges. 

At the end we show that, while RL may in principle allow for model-free learning, reliance on some form of prior knowledge may be in practice required to make the problem solvable. We argue that RL provides an interesting avenue of research for PT not because it allows for pure model-free learning (as in contrast with more traditional model-based artificial intelligence algorithms), but because it may offer a more flexible way to trade off the amount of prior knowledge an agent is provided and the amount of structure an agent is expected to discover. We argue that evaluating this trade-off and implementing agent that take advantage of this would constitute a productive direction of development.

The rest of the paper is organized as follows. Section \ref{sec:Background} offers a review of the main ideas in PT and RL relevant to this work, as well as a review of previous related work. Section \ref{sec:modeling} discusses the problem of modeling PT and CTF as a learning problem, and highlights the specific challenges connected to security. Section \ref{sec:RLModel} gives specific details of our experimental modeling, Section \ref{sec:experiments} presents the results of simulations, and Section \ref{sec:discussion} discusses the results in light of the challenges we uncovered. Section \ref{sec:futurework} suggests future avenues of research, while Section \ref{sec:Ethical} expresses our ethical considerations about this work. Finally, Section \ref{sec:conclusions} summarizes our conclusions.




\section{Background \label{sec:Background}}

In this section we provide the basic concepts and ideas in the fields of PT and RL, and we review previous applications of machine learning to the PT problem. 

\subsection{Penetration Testing \label{sec:Hack}}

Modern computer systems, digital devices and networks may present several types of vulnerabilities, ranging from low-level software binary exploitation to exploitation of network services. These vulnerabilities can be the target of hackers having multiple types of motivations and ways of attacking. PT assumes the perspective of such hackers and performs attacks in order to unveil potential vulnerabilities. 


\subsubsection{Hacking attacks}

Although there is no strict rule on how to carry out a hacking attack, it is still possible to identify the steps that are common in many scenarios.
From the perspective of the attacker, hacking basically consists of steps of \emph{information gathering} 
and steps of \emph{exploitation}. 

%

In the first stage of information gathering an attacker usually collects technical information on the target by probing the system (e.g.: mapping the website content to finding useful information, identifying the input parameters of server side scripts). 
Probing the system in order to acquire relevant information to determine the presence of a specific vulnerability often constitutes the bottleneck in the attack process. Vulnerabilities may be very different, and the second step of exploitation requires understanding the dynamics of the the target system and tailoring the actions to the identified weakness. An attacker has generally to rely on a wide spectrum of competences, from human logic to intuition, from technical expertise to previous experiences. 
After successful exploitation the attacker usually has multiple ways to proceed depending on the aim of the attack. It may simply keep an open unauthorized channel to its target, it may extract private or protected information, or it may use the target system to carry on further attacks.

A notable example of hacking that may be the concern of PT is \emph{web hacking}. The process of web hacking can be decomposed in several successive and alternative steps. Typically, the attacker starts by identifying a target web service through a port scanning and by establishing a communication with the service over the HTTP protocol. She can then access the website files inside the webroot folder, download them, process them within a web browser, and execute locally client-side scripts. Data can also be sent to the remote files in order to be processed on the server-side and obtain customized web responses. Server-side scripts can do many complex actions, such as querying a database, or reading and writing local files; the attacker may send well-crafted inputs in order to compromise these operations. Although web pages may present different and sometimes unique vulnerabilities, typical vulnerabilities can identified and classified \cite{owasp}.




\subsubsection{Capture the flag hacking competitions}
CTFs are a practical learning platform for ethical hackers \cite{CTFTime}. CTF events are normally organized as 48-hour competitions during which different hacking challenges are provided to the participants. 

CTF competitions usually present a set of well-formalized challenges. Each challenge is defined by one vulnerability (or a chain of vulnerabilities) associated with a \emph{flag}. The aim of a participant is to exploit the vulnerability in each challenge, and thus capture the associated flag. No further steps are required from a player (such as, sending data to a command and control server or maintaining the access); the capture of a flag provides an unambiguous criterion to decide whether a challenge was solved or not.
Challenges may be classified according to the type of problem they present (e.g., web hacking challenge or binary exploitation). Normally, human factors are excluded from the solution, so that an attacker has to rely on her knowledge and reasoning, but not on social engineering. In some instances, information about the target system and the vulnerability may be provided to the participants.

Standard CTFs run in \emph{Jeopardy mode}, meaning that all the participants are attackers, and they are presented with a range of different static challenges. In other variants, participants may be subdivided in a \emph{red team}, that is a team focused on attacking a target system, and a \emph{blue team}, that is a team tasked with defending the target system. Alternatively, each team may be provided with an infrastructure they have to protect while, at the same time, attacking the infrastructure of other teams. These last two variants of CTF defines non-static, evolving vulnerabilities, as the defenders in the blue team can change the services at run time by observing the red team actions and patching their own vulnerabilities.

In sum, CTFs, especially in the Jeopardy mode, define a set of well-defined problems that can capture the essence of PT and that can be easily cast in the formalism of games. 

\subsection{Reinforcement Learning \label{sec:RL}}

The reinforcement learning (RL) paradigm offers a flexible framework to model complex control problems and solve them using general-purpose learning algorithms \cite{sutton2018reinforcement}. A RL problem represents the problem of an agent trying to learn an \emph{optimal behavior} or \emph{policy} within a given environment. In model-free learning, the agent is given minimal information about the environment, its dynamics, and the nature or the effects of the actions it can perform; instead, the agent is expected to learn a sensible behavior by interacting with the environment, thus discovering which actions in which states are more rewarding, and finally defining a policy that allows it to achieve its objectives in the best possible way.

\subsubsection{Definition of a RL problem}
Formally, a RL problem \cite{sutton2018reinforcement} is defined by a \emph{tuple} or a \emph{signature}: 
\[
\left\langle \mathcal{S},\mathcal{A},\mathcal{T},\mathcal{R}\right\rangle 
\]
where:
\begin{itemize}
	\item $\mathcal{S}$ is the \emph{state set}, that is the collection of all the states of the given environment;
	\item $\mathcal{A}$ is the \emph{action set}, that is the collection of all the actions available to the agent;
	\item $\mathcal{T}:P\left(s_{t+1}\vert s_{t},a_{t}\right)$ is the \emph{transition function of the environment}, that is the probability for the environment of transitioning from state $s_{t}$ to state $s_{t+1}$ were the agent to take action $a_{t}$;
	\item $\mathcal{R}:P\left(r_{t}\vert s_{t},a_{t}\right)$ is the reward 	function, that is the probability for the agent of receiving reward $r_{t}$ were the agent to take action $a_{t}$ in state $s_{t}$.
\end{itemize}
In this setup, it is assumed that the state of the environment is perfectly known to the agent.
This setup constitute a \emph{fully observable Markov decision process} (MDP) \cite{sutton2018reinforcement}.

The behavior of the agent is encoded in a behavior policy:
$
\pi\left(a_{t}\vert s_{t}\right)=P\left(a_{t}\vert s_{t}\right),
$
that is a probability distribution over the available actions $a_{t}$ given the state $s_{t}$ of the environment. The quality of a policy is measured as its \emph{return}, that is the sum of the expected rewards over a time horizon $T$:
\[
G_{t}^{\pi}=\sum_{t=0}^{T}\gamma^{t}E\left[r_{t}\right],
\]
where $\gamma<1$ is a discount factor that underestimate rewards in the far future with respect to rewards in the near future. The discount factor provides a formal solution to the problem of a potentially infinite sum (for $T\rightarrow \infty$), and an intuitive weighting that makes our agent favor close-in-time rewards instead of postponement. Given this notion of return, the aim of the agent is to learn the optimal policy $\pi^{*}$ that maximizes the return $G$,
that is the policy $\pi^{*}$, not necessarily unique, such that no other policy $\pi$ produces a higher return. Learning an optimal policy requires the agent to balance between the drive for \emph{exploration} (finding previously unseen states and actions that provide high reward) and for \emph{exploitation} (greedily choosing the states and actions that currently are deemed to return the best rewards).

Interaction with the environment (and, therefore, learning) happens over \emph{steps} and \emph{episodes}. A step is an atomic interaction of the agent with the environment: taking a single action $a_{t}$ according to the policy $\pi$, collecting the reward $r_{t}$, and observing the environment evolving from state $s_{t}$ to state $s_{t+1}$. An episode is a collection of steps from an initial state to an ending state.

Notice that during different episodes, even if the signature of the RL problem $\left\langle \mathcal{S},\mathcal{A},\mathcal{T},\mathcal{R}\right\rangle $ is unchanged, the setup may be different. An RL agent is trained not to solve just one specific instance of a problem, but an entire set of problems with a similar structure described by the formalism $\left\langle  \mathcal{S},\mathcal{A},\mathcal{T},\mathcal{R}\right\rangle $. This variability among episodes is important, and it allows a RL agent to \emph{generalize}.  

\subsubsection{Algorithms for RL}

Several algorithms have been proposed to solve the RL problem. 
One of the simplest, yet well-performing, family of RL algorithms is the family of \emph{action-value methods}. These algorithms tackle the problem of learning an optimal policy $\pi^{*}$ through a proxy function meant to estimate the value of each pair of (state, action). 

Formally, these methods define an action-value function for a policy $\pi$ as:
\[
q^{\pi}\left(s_{t},a_{t}\right)=E\left[G_{t}\vert s_{t},a_{t}\right],
\]
that is, the action-value function for a pair (state, action) is the expected return from state $s_{t}$ after taking action $a_{t}$ according to policy $\pi$. Generally, action-value methods follow an approach to learning an optimal policy called \emph{generalized policy iteration} based on two steps: 
\begin{enumerate}
	\item given a starting policy $\pi_{0}$ interact with the environment to learn an approximation of the function $q^{\pi_{0}}\left(s_{t},a_{t}\right)$; 
	\item improve the policy $\pi_{0}$ by defining a new policy $\pi_{1}$
	where, in each state $s_{t}$, the agent takes the action $a_{t}$ that maximizes the action-value function $q^{\pi_{0}}\left(s_{t},a\right)$, that is $\pi^{1}\left(a_{t}\vert s_{t}\right)=\begin{cases}
	1 & \textrm{if }a_{t}=\arg\max_a q^{\pi_{0}}\left(s_{t},a\right)\\
	0 & \textrm{else}
	\end{cases}$. 
\end{enumerate}
Iteratively repeating this process (and allowing space for exploration), the agent will finally converge to the optimal policy $\pi^{*}$. While the second step is quite trivial, consisting just of a maximizing operation, the first step requires fitting the action-value function, and it may be more challenging and time-consuming. There are two main ways of representing the action-value function:
\begin{itemize}
	\item \emph{Tabular representation}: this representation relies on a matrix or a tensor $Q$ to exactly encode each pair of (state, action) and estimate its value; tabular representations are simple, easy to examine, and statistically sound; however they have limited generalization ability and they do not scale well with the dimension of the state space $\mathcal{S}$ and the action space $\mathcal{A}$.
	\item \emph{Approximate representation}: this representation relies on fitting an approximate function $\hat{q}$; usual choices for $\hat{q}$ are parametric functions ranging from simple linear regression to complex deep neural networks; approximate functions solve the problem of dealing with a large state space $\mathcal{S}$ and action space $\mathcal{A}$, and provide generalization capabilities; however they are harder to interpret and they often lack statistical guarantees of convergence.
\end{itemize}

\subsubsection{Q-Learning. \label{ssec:Q-learning}}
A standard action-value algorithm for solving the RL problem is \emph{Q-learning}. Q-learning is a temporal-difference off-policy RL algorithm; \emph{temporal-difference} means that the algorithm estimates the action-value function $q\left(s_{t},a_{t}\right)$ starting from an initial guess (bootstrap), and updates step-by-step its estimation with reference to the value of future states and actions; off-policy means that Q-learning is able to learn an optimal policy $\pi^{*}$ while exploring the environment according to another policy $\pi^{b}$. Q-learning constitutes a versatile algorithm that allows to tackle many RL problems; it can be implemented both with a tabular representation $Q$ of the action-value function or with an approximate representation $\hat{q}\left(s_{t},a_{t}\right)$.

Formally, given a RL problem $\left\langle \mathcal{S},\mathcal{A},\mathcal{T},\mathcal{R}\right\rangle $ with a discount $\gamma$, an agent interacting with the environment in real-time can gradually construct an approximation of the true action-value function $q\left(s_{t},a_{t}\right)$ via a tabular representation by gradually updating its estimation according to the formula:
\begin{equation} \label{eq:QFormula}
Q\left(s_{t},a_{t}\right)\leftarrow Q\left(s_{t},a_{t}\right)+\alpha\left[r_{t}+\gamma\max_{x}Q\left(s_{t+1},x\right)-Q\left(s_{t},a_{t}\right)\right],
\end{equation}
where $\alpha\in\mathbb{R}$ is a scalar defining a step-size \cite{sutton2018reinforcement}. Intuitively, at every step the estimation of $Q\left(s_{t},a_{t}\right)$ moves towards the true action-value function $q\left(s_{t},a_{t}\right)$ by a step $\alpha$ in a gradient ascent-like way.

\subsection{Related Work}

Automated tools for PT consists mainly of security scanners that can send predefined requests and analyze the answers in order to detect specific vulnerabilities (e.g.: Nessus \cite{Nessus}). These tools heavily rely on human knowledge: experts defines scripts that encode the structure of the problem and analyze the collected information. Some applications, such as sqlmap \cite{Sqlmap}, may perform exploitation too, although always with some degree of user interaction. 

Automating the whole process of developing PT strategies has been the object of study for some time, and different models have been proposed to tackle the problem, such as \emph{attack graphs}, \emph{Markov decision process}, \emph{partially observable Markov decision processes} \cite{sarraute2013penetration}, \emph{Stackelberg games} \cite{speicher2019towards}, or \emph{Petri nets} \cite{bland2020machine}. 
Many of the existing solutions follow a \emph{model-based} approach: a PT scenario is first encoded in one of these well-defined models relying on domain expertise, and then processed using model checking or artificial intelligence algorithms to produce optimal plans \cite{boddy2005course,hoffmann2015simulated,applebaum2016intelligent}. Although effective, these models are always limited by the necessity of having human experts defining the dynamics of the models. More recently, the use of model-free RL algorithms has been proposed to tackle PT problems \cite{bland2020machine, ghanem2020reinforcement}. Instead of relying on a model carefully designed by an expert, a model-free agent can interact with an environment by itself and infer an optimal strategy. This line of research has been studied in \cite{pozdniakov2020smart}, with the implementation of tabular and approximate Q-learning algorithms to tackle a paradigmatic PT problem; our work follows the same approach, although our study focuses on a critical assessment of model-free RL agents across a set of prototypical CTF problems, and on the evaluation of different RL techniques aimed at addressing the specific problems we have encountered.

It is also worth mentioning that the encounter between PT and RL has been promoted by DARPA through the Cyber Grand Challenge Event hosted in Las Vegas in 2016 \cite{CyberGrand}. This challenge was a CTF-like competition open to automated agents. The organizers developed a special environment called DECREE (DARPA Experimental Cyber Research Evaluation Environment) where the operating system executed binary files in a modified format and only 7 system calls 
were available. Our work takes inspiration from this challenge, and it aims at studying model-free RL agents that may be deployed to solve similar simplified CTF problems.



\section{Modeling PT as a RL Problem \label{sec:modeling}}

In this section we discuss how we can model PT as a RL problem by examining the challenges and the opportunities in this task. We start by arguing that PT can be naively seen as another game that can be solved by RL. We then move to discuss the specific issue in dealing with PT as a game, that is, the limited access to the structure of the learning problem underlying the CTF challenge. Finally, we present some techniques that may help with this problem by introducing small amounts of a priori knowledge in our model.


\subsection{PT as a Learning Problem}
PT, especially when distilled as a CTF problem, may be easily expressed in terms of a game. It is immediate to identify the players of the game (a red team and a blue team), the rules of the game (the logic of the target system), and the victory condition (capture of the flag). Given the success of RL in tackling and solving games, it seems natural to try to express PT as a game.
Furthermore, at first sight, the distinction between the types of actions performed by an attacker (\emph{information gathering} and \emph{exploitation}) seems to reflect the same division between \emph{exploration} actions and \emph{exploitation} actions in RL. Since RL is assumed to learn to balance exploration and exploitation, it may seem that the PT problem would perfectly fit the RL paradigm.

However, casting the PT problem as a simple game solvable by RL risks missing some challenges peculiar to PT. 
The difficulty for an artificial agent to solve a CTF problem is due several factor. Common challenges are the sheer size of the action and state space which entails a high time and space complexity (in the case of PT, the number of commands an agent can send may be very large) and the limited number of channels through which the agent may acquire information to perform inferences (in the case of PT, an artificial agent may learn only by trial-and-error while a human hacker may rely on alternative sources of knowledge, deductions, hypothesis testing, and social engineering). However, a very specific challenge in solving PT problem follows from the \emph{limited and non-stationary structure} of the problem. 

\subsection{Structure of a Learning Problem}
A RL agent is able to solve a problem by exploiting some \emph{structure} underlying the problem itself. In other words, the problem presents some regularities, patterns, and weakness that the agent may discover and exploit. The structure is captured by an agent in the probability distribution of its policy; as it interacts with the environment, the agent updates its policy and reconstructs the structure of the problem. 

Now, while games tend to have a defined structure expressed in their rules, PT problems may actually expose little structure by obfuscating the logic of the target system. From the perspective of a red team agent, a target system may present different levels of structure. 
At one extreme, we have \emph{perfect systems}, that is systems where defense has no vulnerabilities; these systems are of no interest here, since nothing but failure could be learned either by a human or artificial attacker.
Similarly challenging are \emph{max-entropy systems}, that is systems that have a vulnerability but they have no structure allowing an attacker to find this vulnerability. A max-entropy system is a system where each action $a_{t}$ or set of actions $\left\{ a_{t_i}\right\}$ of the attacker returns as information only whether that action $a_{t}$ or set of actions $\left\{ a_{t_i}\right\}$ was successful or not; no further inference about other actions may be drawn from the feedback. In this setup, if we represent the starting knowledge of the agent as a policy with a uniform distribution over all actions or over all set of actions, then every interaction will provide only a single bit of information, that is the binary outcome of the chosen action or set of actions; no information is provided for the agent to learn about other possible courses of action and thus decrease the entropy (uncertainty) of its policy. A max-entropy system is not absolutely secure, but, provided that there must be a vulnerability, is the safest possible static configuration for a defender. Indeed, the only possible strategy for an attacker against such a system is just to try out all the possible actions. As such, we do not take into consideration this type of problem as the policy or strategy to be learned is structureless and trivial\footnote{This setup may be better suited to be formalized as a \emph{multi-armed bandit problem} \cite{lattimore2018bandit}.}.
We instead focus on \emph{CTF systems}, that is systems that have a vulnerability and have enough structure to allow an attacker to find such a weakness. In this scenario, actions taken by the agent may not lead to success, but they can still leak information useful for the agent to infer the structure of the problem and direct its future actions. This setup is consistent with an actual CTF game, where the red team players, by reasoning and following their intuitions, can discover and exploit the vulnerability. By analogy, an artificial agent is expected to exploit the structure of a system to learn an optimal strategy.

Another radical difference between games and PT revolves around stationarity. While games would normally be defined by a stable structure enforced by a set of unchanging rules, the behaviour of a target system may be radically changed in response to the actions of an attacker. Such a change could be absolutely non-deterministic (e.g., initiated by a blue team player) and it may alter the structure of the game to the point that all the previous exploratory actions of the agents would be rendered useless. When we consider \emph{non-stationary environments}, we need to restrict our attention to systems which are steady enough for the agent to map out and exploit the structure; otherwise, agents that do not take into account this non-stationarity would be bound to fail in learning any meaningful structure.

Environments with limited, and potentially non-stationary, structure constitute a serious problem for model-free RL agents.
From the point of view of the defender, a system to be protected ideally exchanges with the potential attacker messages carrying as little information as possible, and it changes its configuration frequently. In such a setting, the biggest challenge for a red team RL agent is not to learn on optimal strategy over a known structure (as in the case of traditional games), but to discover efficiently the structure of the system itself. Thus, CTFs stress the need for \emph{exploration}: for an RL agent managing an efficient exploration is as important as developing a complex exploitation strategy. Good RL algorithms for PT should take this aspect in particular consideration.

\subsection{A Priori Knowledge of the Structure of a Problem}
Although a RL agent may in principle learn structure from scratch in a pure model-free way, this may turn out to be a computationally hard challenge: methods based on the explicit enumeration of (state, action) pairs like tabular Q-learning face a combinatorial explosion of (state, action) pairs to evaluate; even methods based on parametric approximation like neural Q-learning may need to sample and explore a large space. Injecting some form of elementary a priori knowledge about the structure of the problem may greatly simplify the learning problem. Some basic forms of a priori knowledge are:
\begin{itemize}
    \item \emph{Lazy loading:} this is an empirical method used with tabular Q-learning consisting in initializing new (state, action) pairs only when the agent encounters them and evaluating them on-the-fly. The underlying assumption of this technique is that the structure of the problem is \emph{sparse}: most of the possible (state, action) pairs will never be encountered, either because logically impossible or inconsistent. Lazy loading may drastically reduce the complexity of tabular Q-learning and make it feasible.
    \item \emph{State aggregation:} this is a technique used to aggregate together (state, action) pairs that are logically identical. This corresponds to stating that the structure of the problem presents \emph{equivalence classes} over its states, and that the agent may treat these states as equivalent. State aggregation may allow better generalization both in tabular and neural Q-learning.
    \item \emph{Imitation learning} (or \emph{learning from demonstrations} \cite{abbeel2004apprenticeship}): this is a technique used to prime the policy of the agent by providing it with samples of high-performing human or artificial behaviour. This is a \emph{data-driven} approach where the structure of the problem is implicitly communicated to the agent through examples; instead of starting with a blank slate knowledge of state and actions, the agents is provided with examples of series of actions that may lead to a solution. Imitation learning may reduce the learning time of tabular and neural Q-learning algorithms.
\end{itemize}
These techniques may enrich the RL agent, and make learning more efficient. Such an enriched agent goes, to a certain measure, against the model-free paradigm, in the sense that it relies on a degree of expertise to be hard-coded in its algorithm; for instance, state aggregation requires knowledge of which states may considered equivalent, while imitation learning assumes the possibility of collecting or generating successful samples of behaviour. In general, though, all these techniques require less expertise and time that the definition of an explicit model fully describing the logic of the target system. For this reason, although a model relying on these methods is not purely model-free anymore, it is not usually considered model-based.\\

In light of the challenges described above, we will aim to show how crucial the role of structure is by showing that (a) in the limit of a max-entropy system or a highly non-stationary PT environment, a RL agent is bound to learn trivial policies reducing to pure guessing; (b) priming an agent with a measure of a priori knowledge may sensibly reduce the complexity of the problem. These challenges will provide a criterion to study the application of RL to PT and CTF, evaluate our simulations, and suggest future developments.

\section{Formalization of a PT Problem} \label{sec:RLModel}

In this section we move on to propose a formalization of CTF challenges using the formalism of RL. 
We first identify the classes of CTF challenges that we will study experimentally, and then present a precise formalization of these CTF problems using the standards of RL. 


\subsection{Types of CTF Problems} 
CTFs may be categorized in groups according to the type of vulnerability they instantiate and the type of exploitation that a player is expected to perform. Each class of CTF problems may exhibit peculiar forms of structure and may be modeled independently.
In this paper, we will consider the following prototypical classes of CTF problems:
\begin{itemize}
	\item \emph{Port scanning and intrusion}: in this CTF problem, a target server system exposes on the network a set of ports, and an attacker is required to check them, determine a vulnerable one, and obtain the flag beyond the vulnerable port using a known exploit;
	
	\item \emph{Server hacking}: in this CTF problem, a target server system exposes on the network a set of services, and an attacker is required to interact with them, discover a vulnerability, either in the form of a simple unparameterized vulnerability or as a parametrized vulnerability, and obtain the flag by exploiting the discovered vulnerability;
	
	\item \emph{Website hacking}: a sub-type of server hacking, in this CTF problem a target server system exposes on the network a web site, and an attacker is required to check the available pages, evaluate whether any contains a vulnerability, and obtain the flag behind one of the pages by exploiting the discovered vulnerability.
\end{itemize}
These three classes provide well-known tasks that can be modeled as RL problem at various levels of simplification and abstraction.

\subsection{RL Formalism}
We consider as a \emph{RL agent} an artificial red team hacking player interacting with a vulnerable target system. The target system constitutes the \emph{environment} with which the agent interact. The \emph{goal} of the agent is to capture the flag in the target environment in the fastest possible way.
Given the RL problem $\left\langle \mathcal{S},\mathcal{A},\mathcal{T},\mathcal{R}\right\rangle$ we set the following requirements and conditions:
\begin{itemize} 
	\item The state space $\mathcal{S}$ is assumed to be an unstructured finite set of states that encode the state of the environment and, implicitly, the state of knowledge of the agent.
	\item The action space $\mathcal{A}$ is assumed to be an unstructured finite set containing all the possible actions that may be performed by the agent. Notice that the set is the same in any state; even if some action may not be available to the agent in some states, this information is not provided to the agent; an agent is expected to learn by experience which action are possible in any state.
	\item The transition function $\mathcal{T}$	is assumed to be a \emph{deterministic} function that encodes the logic of the specific CTF scenario that will be considered.	
	\item The reward function $\mathcal{R}$ is assumed to be a \emph{deterministic} function defining how well the agent is performing. Rewards will normally be dense but not highly informative: the agent receives a small negative reward for each attempt performed (normally $-1$), and a large positive reward for achieving its objective (normally $100$); this setup will push the agent to learn the most efficient strategy (in terms of attempts) to capture a flag.
\end{itemize}

In the following experimental analysis we will focus on one particular algorithm, that is, \emph{tabular Q-learning}. Our choice is motivated by several factors: (i) in general, Q-learning is a classical and well-performing algorithms, allowing us to relate our results with the literature; (ii) it guarantees that the agent will converge to an optimal policy; (iii) the use of a tabular representation allows for a simpler interpretation of the results; (iv) Q-learning is step-wise fast and efficient, thus allowing us to easily repeat experiments and guaranteeing reproducibility; (v) Q-learning has few hyper-parameters, allowing for a more effective tuning.
The main drawback of adopting tabular Q-learning is scalability, which, implicitly, reduces the complexity of the problems that we will be able to consider. Despite this limitation, though, our results will probe and validate the possibility of solving CTF problems using RL, and they will allow us to assess the relevance of the challenges we identified. 

\section{Experimental Analysis \label{sec:experiments}}

In this section we provide concrete instances of simple CTF challenges, we model them in the form of RL problems using the formalism discussed in Section \ref{sec:RLModel}, and we solve them using Q-learning. We consider CTF challenges with increasing complexity, and as we face the challenges we identified in Section \ref{sec:modeling}, we evaluate the different methods for introducing a priori knowledge that we have reviewed in Section \ref{sec:modeling}.
All the simulations are implemented following the standard RL interface defined in the OpenAI gym library\footnote{\url{https://gym.openai.com/}} \cite{brockman2016openai}, and they are made available online \footnote{ \url{https://github.com/FMZennaro/CTF-RL}} to guarantee reproducibility and further experiments and extensions. Detailed explanations about the action set and hyperparameter configuration of each simulation are provided in the Supplemental Material.

\subsection{Simulation 1: Port Scanning CTF Problem}
In this simulation we consider a very simple \emph{port scanning} problem. We use the basic tabular Q-learning algorithm to solve it, and we analyze our results in terms of structure of the solution and inference steps to convergence. This basic simulation is aimed at showing the capabilities of RL agents in a well-behaved (limited structure available and stationary) scenario and their dependence on the size of the problem.

\textbf{CTF scenario.}
The target system is a server which runs only one service affected by a known vulnerability. The port number on which the service runs is unknown; however, once the service port is discovered, the agent knows for certain where the vulnerable service is and how to exploit it. 
The red team agent can interact with the server by running a port scan or by sending the known exploit to a specific port. 
In this simplified scenario the vulnerability can be targeted with a ready exploit with no parameters; also it is assumed that no actions are performed by the blue team on the target system.     

\textbf{RL Setup.}
We define a target server exposing $N$ ports, each one providing a different service; one of the services is affected by a vulnerability, and behind it lies the objective flag.

We model the action set $\mathcal{A}$ as a collection of $N+1$ actions: one port scan action, and one exploitation action for each of the $N$ existing ports. 
We also model the state set $\mathcal{S}$ as a collection of $N+1$ binary variables: one initial state representing the state of complete ignorance of the agent, and one state for each port taking value of one when we discover it is the vulnerable port.
The dimensionality of a tabular action-value matrix $Q$ scales as $O(N^2)$.


This simple exercise allows us to have a basic assessment of the learning ability of the agent. Notice that the agent is not meant to learn simply the solution to a single instance of this CTF game; in other words, it is not learning that the flag will always be behind port $k$. In every instance of the CTF game the flag is placed behind a different port; thus, the agent has to learn a \emph{generic strategy} that allows it to solve the problem independently from the initial setup.

In general terms, this problem constitutes a very simple challenge, in which the optimal strategy is easily acknowledged to be a two-step policy of scanning and then targeting the vulnerable port with an exploit. However, the RL agent is not aware of the semantics of the available actions and it can not reason out an optimal strategy, but it can only learn by trial and error. 

\textbf{Results.}
We run our simulation setting $N=64$ ports. We randomly initialize the policy of the agent and we run $1000$ episodes. We repeat each simulation $100$ times in order to collect reliable statistics.

\begin{figure}[!t]
	\begin{subfigure}[b]{0.2\textwidth}
		\includegraphics[scale=.3]{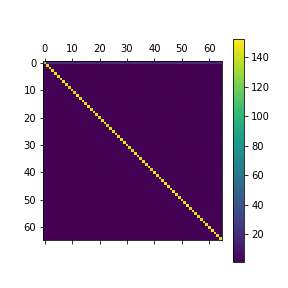}
		\caption{}
	\end{subfigure}
	\begin{subfigure}[b]{0.25\textwidth}
		\includegraphics[scale=.35]{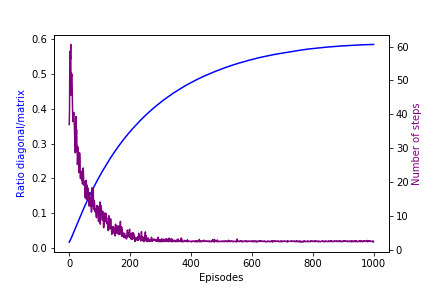}
		\caption{}
	\end{subfigure}
		
	\caption{Results of Simulation 1. (a) Learned action-value matrix $Q$. (b) Plot of ratio $\frac{\sum_{i=0}^{N}Q_{ii}}{\sum_{i,j=0}^{N}Q_{ij}}$ as a function of the number of episodes (in blue), and number of steps as a function of the number of episodes (in purple). \label{fig:simul1}}
\end{figure}

As discussed, in this simple scenario we know what would be the optimal policy and, therefore, what we expect the agent to learn. 
Figure \ref{fig:simul1}(a) shows a plot of the action-value matrix $Q$ at the end of the $1000$ episodes. The matrix shows a clear diagonal pattern, meaning that in state $s_i$, for $0 \leq i \leq N$, the agent has learned to favor action $a_i$. This makes sense: in the initial complete-ignorance state $s_0$ the agent selects action $a_0$ corresponding to the port scan action; in state $s_i$, for $1 \leq i \leq N$, corresponding to the knowledge that port $i$ is vulnerable, the agent selects action $a_i$, corresponding to an exploit on the relative port. We can thus conclude that the agent has successfully learned the desired optimal strategy.
The blue plot in Figure \ref{fig:simul1}(b) shows the convergence towards the optimal strategy as a function of the number of episodes. The y-axis reports the ratio between the sum of the diagonal of $Q$, and the sum of all the entries of $Q$, that is $\frac{\sum_{i=0}^{N}Q_{ii}}{\sum_{i,j=0}^{N}Q_{ij}}$. Since we know that the optimal strategy is encoded along the main diagonal of $Q$, this statistics tells us how much of the mass of $Q$ is distributed along the diagonal. After around $400$ episodes the learning of the agent enters a phase of saturation. Notice that this ratio would converge to $1$ only in an infinite horizon.
The purple plot in Figure \ref{fig:simul1}(b) illustrates the number of steps per episode. After around $200$ episodes the agent has learned the optimal strategy and completes the challenges in the minimum number of actions.

\textbf{Discussion.}
The success of RL in this proof-of-concept simulation is not surprising; yet, it highlights the specific challenges of addressing hacking using RL: solving the CTF challenge requires learning the structure of the problem; this is feasible, but, using only experiential data and inference 
means that the RL agent has to rely strongly on exploration. Almost two hundred episodes were necessary to converge to a solution, a number of attempts far greater than what necessary for a human red team to find an optimal strategy. 

\subsection{Simulation 2: Non-stationary Port-scanning CTF Problem}
In this simulation we extend the previous problem by considering a more challenging scenario in which the target system is not stationary, but it may randomly change in response to the actions of the agent. 

\textbf{CTF scenario.}
In this scenario the blue team is not passive anymore, but it can act in response to actions perpetrated by the red team. We setup the same target system as before: the server has a single exploitable service running on a port whose number is unknown to the attacker. To model an attack-defense scenario, we suppose that the blue team is aware of the exploitable service but that they cannot stop it because this would affect their continuous business operation. The blue team cannot filter out traffic, and the only option they have is to move the service to another port if they observe actions that may prelude to an attack. This case is rather unrealistic, but we use it as a simplified attack-defence contest with limited actions. 

\textbf{RL Setup.}
We consider the same port scanning scenario defined in the previous simulation. However, we add a non-stationary dynamic: whenever the attacker uses a port scan action, the target server detects it with probability $p$; if the detection is successful the flag is randomly re-positioned behind a new port.
Given the non-stationarity, this problem constitutes a more challenging learning problem than the previous one. In particular, knowledge of the structure gained by the agent via port scanning may not be reliable. In this stochastic setting, the optimal strategy is not necessarily the deterministic policy used in Simulation 1.

\textbf{Results.}
We run our simulation setting $N=16$ ports. All the remaining parameters of this simulation are the same as in Simulation 1. We consider all the possible values of $p$ in the set $\{ 0, 0.1, 0.2, ... , 0.9, 1.0 \}$. We repeat each simulation $100$ times in order to collect reliable statistics.

\begin{figure}
	\begin{subfigure}[b]{0.15\textwidth}
		\includegraphics[scale=.3]{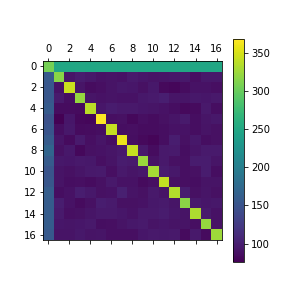}
		\caption{}
	\end{subfigure}
	\begin{subfigure}[b]{0.15\textwidth}
		\includegraphics[scale=.3]{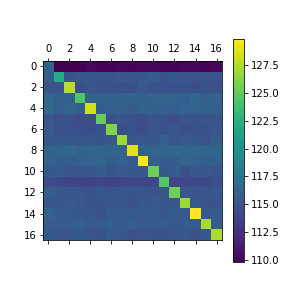}
		\caption{}
	\end{subfigure}
	\begin{subfigure}[b]{0.15\textwidth}
		\includegraphics[scale=.3]{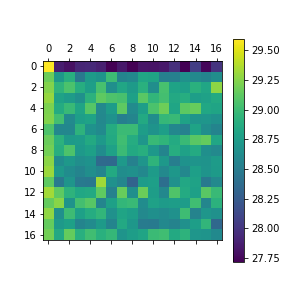}
		\caption{}
	\end{subfigure}
	
	\caption{Results of Simulation 2. Learned action-value matrix $Q$ for: (a) $p=0.1$, (b) $p=0.5$, and (c) $p=1$. \label{fig:simul2_matrices}}
\end{figure}

\begin{figure}
	\begin{center}		
		\includegraphics[scale=.4]{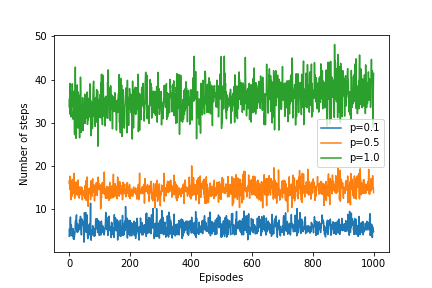}
	\end{center}
	
	\caption{Results of Simulation 2. Number of steps as a function of episodes for $p=0.1$, $p=0.5$, $p=1$. \label{fig:simul2_steps}}
\end{figure}

Figure \ref{fig:simul2_matrices} reports the action-value matrices learned for $p=0.1$, $p=0.5$ and $p=1$. While for small value of $p$ the action-value matrix $Q$ resembles closely the pattern we observed in Simulation 1, for higher values of $p$ we lose this structure. In the almost-deterministic case $p=0.1$ (Figure \ref{fig:simul2_matrices}(a)) it is reasonable to use a port scan action at the beginning, followed by an exploit action that has a high probability of success; therefore we observe the usual diagonal shape. In the more stochastic case $p=0.5$ (Figure \ref{fig:simul2_matrices}(b)) it is likely that a port scan action is detected and that the flag is moved; yet using a port scanning action and a targeted action is still a reasonable bet, even if less effective (notice the different scale for the matrices in Figure \ref{fig:simul2_matrices}(a) and Figure \ref{fig:simul2_matrices}(b)). Finally in the completely random case $p=1$ (Figure \ref{fig:simul2_matrices}(c)) a port scan action certainly results in a detection, and no plan can be built over the information gathered; the agent is basically reduced to resort to plain random guessing.
Consistently, Figure \ref{fig:simul2_steps} shows the number of steps per episode when using $p=0.1$, $p=0.5$, $p=1$. In the almost-deterministic case, the number of episodes sets almost immediately close to optimal; as we increase the stochasticity the number of steps increases because the agent can only try to guess the location of the vulnerability. Notice that the average number of steps in the completely random setting is higher than the number of ports; this is due to the fact that the agent tries out from time to time the port scan action, thus causing the flag to move, and requiring the agent to re-try its exploit on already checked ports.

\textbf{Discussion.} The introduction of a non-stationary dynamics makes the problem more challenging, by preventing the agent to learn the exact structure of the problem with certainty. Despite this, thanks to its formalization, a Q-learning agent is still able to solve this CTF problem in a reasonable, yet sub-optimal, way, as allowed by the degree of stochasticity and non-stationarity.
For $p=0$, the CTF system has a clear structure and it can learn an optimal policy; as $p$ increases, we slowly move from what we defined as a \emph{CTF system} to a \emph{max-entropy system} (see Section \ref{sec:modeling}). Indeed, for $p=1$ our problem represents a \emph{max-entropy system}: no action provides actual information on the structure of the target server (the port scan action is essentially unreliable and useless); unable to reconstruct any structure, RL has a very limited use: all we can do is just guessing, that is trying out one by one all the ports looking for the vulnerability. This underlies the role of structure in learning using RL agents.

\subsection{Simulation 3: Server Hacking CTF Problem with Lazy Loading}
In this simulation we consider a more realistic problem representing a simple server hacking scenario. Although very simplified, this problem already presents a serious challenge to the tabular Q-learning agent because of the size of its Q-table. To solve this obstacle we rely on a priori knowledge in the form of lazy loading, thus controlling the dimensionality of the state and action space and pruning non-relevant states. We analyze how learning happens under this scenario, and what is the effect of the adopted approach on inference.

\textbf{CTF scenario.}
In this simulation a target server provides different standard services, such as web, FTP, or SSH. Each service may have a vulnerability, either a simple vulnerability easily exploitable without a parameter (such as a Wordpress page with a plugin that may lead to an information disclosure in a specific known URL) or a vulnerability requiring the attacker to send a special input (such as a Wordpress plugin with SQL injection).

The attacker can carry out three types of information gathering actions. (i) It can check for open ports and services on the server. (ii) It can try to interact with the services using well-known protocols; this allows it to obtain basic information (such as banner information), and discover known vulnerabilities, such as weaknesses recorded in a vulnerability databases. (iii) It can interact more closely with potentially unique service setups or customized web pages; this will allow the attacker to identify undocumented vulnerabilities and the input parameters necessary for exploitation; for instance, in case of a FTP service, the agent may discover the input parameters for username and password, or, in the case of more complex services such as web, it may obtain GET and POST web parameters. 
In addition, the attacker has also two exploitation actions. (i) It can exploit a non-parametrized vulnerability by accessing the vulnerable service and retrieving the flag. (ii) It can choose a parameter out of a finite pre-defined set, and send it to a service to exploit a parametrized vulnerability and obtain the flag.
In this scenario we make the simplified assumption that the agent can identify just a parameter name from a fixed and limited set, and it does not need to select a parameter value.


\textbf{RL Setup.} 
We define a target server exposing $N$ ports, each providing one of $V$ different services. One of the services is taken to be flawed, and behind it lies the objective flag. The vulnerability may be a simple \emph{non-parametrized vulnerability} or a \emph{parametrized vulnerability}. In the last case, the vulnerability may be \emph{already known}, or it may be \emph{previously unknown} thus requiring deeper probing and analysis of the service. The parameter for the parametrized vulnerability is chosen out of a set of $M$ possible parameters.

The collection of basic actions available to the agent gives rise to a larger set $\mathcal{A}$ of concrete actions, where each action type is instantiated against a specific port. The set of states $\mathcal{S}$ has a large dimensionality as well, due to the problem of tracking what the agent has learned during its interaction with the server.
As a rough estimation, in our implementation we estimate the number of total states as:
\begin{equation} \label{eq:nStates}
\left| \mathcal{S} \right| \approx 2^{11}N^{3}VM.
\end{equation}
Refer to the Supplemental Material for the derivation of this approximation. The encoding used to track the state forms a sufficient statistics that tracks all the actions of the agent and records all its knowledge. It is not meant to be an optimal encoding, and the dimensionality of the set $\mathcal{S}$ may be reduced through a smarter representations of the states. However, even if we were to make the encoding more efficient, the overall dimensionality would quickly become unmanageable when the parameters $N$, $M$ or $V$ were to grow.
A pure tabular Q-learning agent would require a prohibitively large amount of memory just to instantiate its Q-table. However, just relying on the simple knowledge that several (state, action) pairs that may not be relevant or informative, we adopt a \emph{lazy-loading} approach: instead of instantiating from the start a large unmanageable action-value matrix $Q$, we progressively build up the data structure of $Q$ as the agent experiences new (state, action) pairs.




This problem constitutes a more realistic model of a CTF challenge, presenting a target system with multiple services, each one potentially having different types of weaknesses (unparameterized and parametrized vulnerabilities) at different levels (easy vulnerabilities already known or more treacherous vulnerabilities yet unknown). In this more challenging problem it is harder to define a simple deterministic optimal solution as it was in Simulation 1. A standard approach is undoubtedly to use more exploratory actions at the beginning, and leave exploitative actions for the end. However, the variability in the location of the flag and the sharp dynamics of the system make the problem far from trivial.


\textbf{Results.}
We run our simulation setting $N=4$ ports, $V=5$ services, $M=4$ parameters. We randomly initialize pairs of (state, action) at run-time, and we run the agent for $10^6$ episodes. We repeat each simulation $20$ times in order to collect reliable statistics in a feasible amount of time.

\begin{figure}
	\begin{subfigure}[b]{0.23\textwidth}
		\includegraphics[scale=.3]{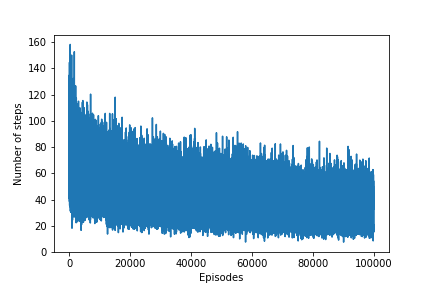}
		\caption{}
	\end{subfigure}
	\begin{subfigure}[b]{0.2\textwidth}
		\includegraphics[scale=.3]{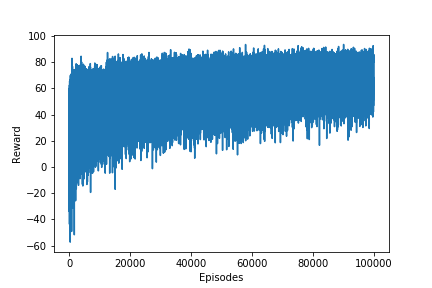}
		\caption{}
	\end{subfigure}
	
	\caption{Results of Simulation 3. (a) Plot of number of steps as a function of the number of episodes; (b) plot of reward as a function of the number of episodes. \label{fig:simul3_1}}
\end{figure}

\begin{figure}
	\begin{center}		
		\includegraphics[scale=.3]{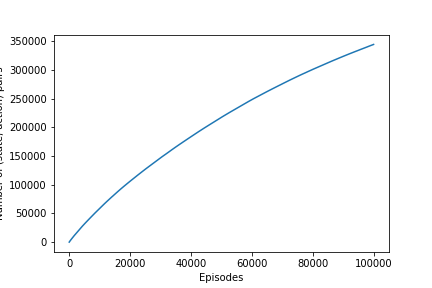}
	\end{center}
	
	\caption{Results of Simulation 3. Number of entries in the action-value table $Q$ as a function of number of episodes. \label{fig:simul3_2}}
\end{figure}

Figure \ref{fig:simul3_1}(a) reports the number of steps taken by our agent to complete a task, and, conversely \ref{fig:simul3_1}(b) shows the reward obtained by the agent. These plots are quite noisy, but they show a clear improvement in the first few thousand episodes: we can clearly see a drop in the number of steps and an increase in the amount of reward collected. Notice that the high variance recorded is in part due to the highly exploratory behavior of the agent ($\epsilon=0.3$) that leads the agent to take a random action almost one third of the times. Interestingly, though, the upper bound of the reward curve approaches a reward of $80$ or higher, pointing out that the agent was indeed able to learn a sensible strategy as it was able to solve the CTF problem in few actions compared to the large number of possible combinations of actions it could try.
Figure \ref{fig:simul3_2} shows the number of entries in the action-value table $Q$ during the episodes. The plot seems to have a parabolic behavior growing fast at the beginning and slowing down towards the end. This makes sense, as at the beginning every state encountered by the agent is new and needs to be added to the table $Q$. The continual increase in size is due to the strong exploratory policy ($\epsilon=0.3$) followed by the agent. Notice, that if we were to substitute the values of $N$, $V$ and $M$ of this simulation in Equation \ref{eq:nStates} we would get a rough estimate for $\left|S\right|$ of over $2\cdot 10^6$; therefore the number of states learned so far is an order of magnitude smaller ($3.5 \cdot 10^5$), and it has allowed the agent to learn swiftly a reasonable policy with a significantly smaller consumption of memory. 

\textbf{Discussion.}
This more realistic simulations highlights at the same time the standard strengths and weaknesses of RL agents. An RL agent may be able to tackle a challenging problem with a subtle and sharp structure like the one presented, but, potentially, at a high computational cost. A trivial implementation may still be able to solve the problem, but it may quickly become unmanageable if it were to treat explicitly all the possible states. Instead of relying on the knowledge that some states are important and others are not, lazy loading has allowed the agent to discriminate between relevant and non-relevant states based on its experience.

\subsection{Simulation 4: Website Hacking CTF Problem with State Aggregation}
Lazy loading is a simple technique to inject a priori knowledge that may allow an agent to learn effectively in moderately complex environment by neglecting non-relevant (state, action) pairs. However the usefulness of lazy loading is limited, since, within inherently complex environments with a large amount of relevant states, the number of (state, action) pairs to be recorded may be unmanageable. In this simulation we run an environment similar to the previous one, but we adopt the additional strategy of performing \emph{state aggregation} over similar states. Again, we run our simulations and we study the dynamics and the performance of inference and learning.

\textbf{CTF scenario.}
In this simulation we assume that the attacker knows the location of a target web page, so no port scan or protocol identification is required. The webpage consists of a set of files: starting from an index file, the attacker can map the visible files by reading the HTML content and by following the links inside the content. The webpage may also host hidden files not linked to the index. Some of the files contain server-side scripts and the attacker may identify customized inputs that may be sent to perform an exploitation and capture the flag.
The attacker is given three types of information gathering actions. (i) It can read the index file, follow recursively all links, and thus obtain a map of all the linked files on the server. 
(ii) It can try to find hidden files by parsing the content of a visible file and infer the existence of hidden files; for instance, looking at a file on a Wordpress site, the attacker may suspect the existence of \emph{/wp-login/index.php}. (iii) It can analyze a visible or hidden file in order to find input parameters that can be used for an exploitation. A single exploitation action is possible. (i) The attacker can send an input parameter to a file and, if correctly targeting the vulnerable file, obtain the flag. Here, again we restrict our model to the problem of identifying a vulnerable parameter name out of a set, and not its parameter value. 

\textbf{RL Setup.}
We define a target server hosting $N$ files, partitioned in $N_{vis}$ visible files and $N_{hid}$ hidden files. Visible files are linked to the index file and connected among them in a complete graph; hidden files are files not openly linked to the index files but referenced or related to one of the visible files. One of the files, either visible or hidden, contains a parametrized vulnerability behind which lies a flag. The vulnerable parameter is chosen out of a set of $M$ possible parameters.

As before, the dimensionality of the action-value matrix grows exponentially with the number of files $N$ and the number of parameters $M$. In order to make the problem manageable we introduce a degree of prior knowledge in our model. We know that files on the target servers may be different, but the way to interact with them is uniform: we explore and inspect files using the same actions; we target files with the same vulnerability in an identical way. Notice that, in the real-world, the concrete way in which we implement actions on different files may be different, but these distinctions are abstracted away in the current model. The dynamics of interacting with files are then homogeneous among all the files. Thus, instead of requiring the agent to learn a specific strategy on each file, we instruct it to learn a single policy that will be used on all the files. We achieve this simplification using \emph{state aggregation} \cite{sutton2018reinforcement}. At each time step, the agent will be focused only on a single file, interact with it and update a global policy valid for any file. 



\textbf{Results.} We run our simulations randomly setting $2 \leq N_{vis} \leq 4$ visible files and $0 \leq N_{hid} \leq 2$ hidden files. We randomly initialize pairs of (state, action) using lazy loading and state aggregation. We run a single agent for $10^5$ episodes and then we test it on $100$ episodes during which we set the exploration parameter $\epsilon$ to 0. We repeat the testing of a trained agent $100$ times in order to collect reliable statistics.

\begin{figure}
	\begin{subfigure}[b]{0.23\textwidth}
		\includegraphics[scale=.3]{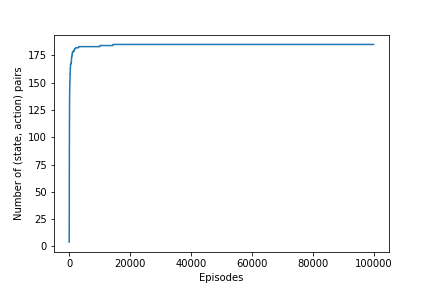}
		\caption{}
	\end{subfigure}
	\begin{subfigure}[b]{0.2\textwidth}
		\includegraphics[scale=.3]{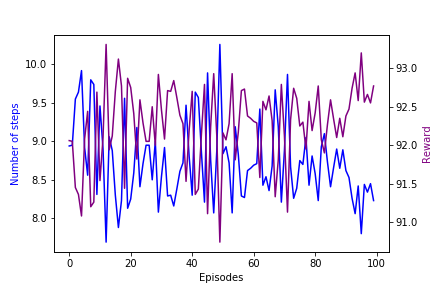}
		\caption{}
	\end{subfigure}
	
	\caption{Results of Simulation 4. (a) Number of entries in the action-value matrix $Q$ as a function of the number of episodes; (b) reward and number of steps as a function of the number of episodes. \label{fig:simul4}}
\end{figure}

Figure \ref{fig:simul4}(a) shows the number of (state, action) pairs in the action-value table $Q$ of our agent during the $10^5$ episodes of learning. The number of states saturates very quickly, enumerating all the $\sim 180$ states encountered by the agent.
Figure \ref{fig:simul4}(b) shows the reward and the number of steps on further $100$ episodes when running the same agent with the exploration parameter ($\epsilon$) set to zero. As expected the two plots are perfectly complementary, with the number of steps oscillating between $8$ and $10$, and the reward between $91$ and $93$. Removing the exploration parameter is a risky choice that may lead the agent to get stuck if it were to face sudden changes in the environment, but it allows us to better appreciate the fact that the agent indeed was able to learn a clear policy that allowed it to capture a flag with a minimal number of actions; eight to ten steps is indeed what is necessary to probe the target server, collect information on the files, and finally retrieve the flag.

\textbf{Discussion.}
This simulation preserves most of the complexity of Simulation 3, and it shows how using proper RL algorithms and techniques (lazy loading and state aggregation), a RL agent may manage to solve effectively a challenging CTF problem. 
Notice that state aggregation allowed us to introduce a form of knowledge that a RL agent would not normally have. A human red team player may reach the conclusion that it is reasonable to act in a uniform way with different files from her previous experience with files; this knowledge provides her with an effective shortcut to reach a solution. A RL agent has no similar possibility as it has no formal concept of files; it could end up learning by inference a policy that is actually uniform for all the files, but this would require collecting a large sample of experiences. State aggregation allowed to inject useful prior information about the structure of the problem, thus simplifying exploration and  reducing the number of (state, action) pairs.

Another interesting feature of this simulation is the use of a graph to represent the filesystem on the target website. 
In this simulation, given the small size of the graph comprising between two and six files we relied on a simple linear exploration of the graph; however, smarter and more sophisticated way of manipulating and exploring the graph may be taken into consideration to exploit the knowledge of this structure and improve the performance of the agent. 

\subsection{Simulation 5: Web Hacking CTF Problem with Imitation Learning}
Lazy loading and state aggregation improve the performance of the agent by reducing the size of the state space; however, if the agent has to explore a large state space and the optimal (or satisfactory) solution occupies a small volume of this space, the search process may take unreasonably long. In this simulation we consider a way to direct the learning process more explicitly by using \emph{imitation learning}, which emulates learning in a teacher-and-student setting, where expert paradigmatic behaviours are offered to a student to speed up its learning. We analyze the behavior of the agent under this setup and we compare the results of this process with the results obtained in the previous simulations.

\textbf{Hacking scenario.} We consider again the same server hacking problem presented in Simulation 3, as this constitutes the most challenging problem we have faced so far.

\textbf{RL Setup.} We consider the same setup used in Simulation 3.
Beyond lazy-loading, this time we also rely on another standard RL technique, that is, imitation learning.
In imitation learning, an agent is provided with a set of $D$ trajectories defined by human experts; in our case, these trajectories encode the behavior of a hypothetical human red team player trying to solve the web hacking CTF problem. These trajectories represent samples of successful behavior and provide information to the RL agent about the relevance of different options. Indeed, in imitation learning, the agent, instead of starting in a state of complete ignorance, is offered examples of how actions can be combined to reach a solution of the problem. This simplifies the exploration problem: instead of searching uniformly in the whole space of policies, the search is biased towards expert-defined policies. This bias allows to solve the problem more efficiently, but it also makes less likely that the agent will discover policies that are substantially different from human behavior.

\textbf{Results.} We run our simulations using the same setting used in Simulation 3. First, we train a standard RL agent for $10^5$ episodes. Then, we train three imitation learning agent, each one being provided with $100$, $200$ and $500$ demonstrations respectively; after that the three imitation learning agents are further trained for $100$ episodes.

\begin{figure}
	\begin{center}		
		\includegraphics[scale=.3]{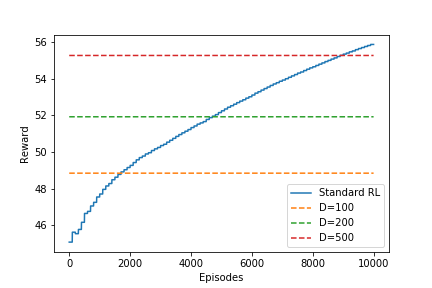}
	\end{center}
	
	\caption{Results of Simulation 5. Reward achieved by RL agents with and without imitation learning (see text for explanation). \label{fig:simul5}}
\end{figure}

Figure \ref{fig:simul5} shows the rewards obtained by the different agents. The dotted lines represent the average reward obtained by the imitation learning agent during the $100$ episodes of training; notice that these lines are independent from the scale on the x-axis and are plotted as constants for reference. The blue line shows the reward, averaged every $100$ episodes, obtained by the standard RL agent during training. The graph shows that the standard RL agent needs to be trained on almost $2000$ episodes before reaching the average reward that an imitation learning agent can achieve with $100$ demonstrations; similarly, the whole training time of $10^5$ episodes is necessary to match an imitation learning agent provided with $500$ demonstrations. The overall rewards are still far from being optimal, but imitation learning allows for a reduction of the number of episodes of training of one order of magnitude.

\textbf{Discussion.}
Imitation learning proved to be an effective techniques to enable faster learning for the RL agent. This improvement is again due to the possibility of introducing in the agent knowledge on the structure of the problem. Indeed, demonstrations are an implicit way to express human knowledge about the structure of the CTF problem: instead of encoding knowledge on the structure of the problem in a formal mathematical way, we provide the RL agent with concrete observations about the structure of the problem. This information can successfully be exploited by the agent in order to learn an optimal policy.

\section{Discussion \label{sec:discussion}}
The simulations in this paper showed the feasibility of using RL for solving CTF problems, as well as the central role that the challenges of discovering structure and providing prior knowledge play in this context. While RL was able to solve optimally a simple CTF with an elementary structure (Simulation 1), we observed that changes in the structure of the CTF problem may make the problem harder to solve. We considered two ways in which the structure of the problem may change and raise concrete challenges. First, a progressively more undefined problem structure, shifting from a stationary CTF system to a max-entropy system, highlighted the limits of learning by inference (Simulation 2). Second, a stationary CTF problem with a progressively more complex structure required an exponential number of samples for the agent to work out the structure of the problem. In this last case, we showed how RL techniques, such as lazy loading, state aggregation, or imitation learning, may allow the RL agent to tackle more complex problems (Simulation 3, 4, 5). These techniques were explained and justified in terms of providing the agent with elementary prior information about the structure of the problem. Lazy loading corresponded to the assumption that certain configurations in the problem space would never be experienced, and therefore could be ignored; state aggregation expressed the assumption that certain configurations would be pragmatically identical to others; and imitation learning codified the assumptions that an optimal solution would not be too far from well-known demonstrations. Notice that while imitation learning necessarily require expert knowledge, lazy loading and state aggregation are based on simple assumptions needing limited expertise. Although implemented in specific simulations, all these forms of prior knowledge are not semantically-tied to a specific problem, and they may be easily deployed across a wide range of other CTF problems. 
Discovering structure is an essential step in solving CTF problem in which an attacker aims at uncovering and exploiting vulnerabilities. Our observations highlighted and remarked this point, leading to the conclusion that ideal learning agent would properly negotiate between discovering structure autonomously and relying on a priori knowledge. Although fully model-free RL agents have the theoretical possibility of discovering relevant structure, complexity considerations and practical limitations suggest that the introduction of a priori knowledge may be desirable. While fully model-based agents may not be ideal because hard to encode and not very versatile, RL agents that combine model-free algorithms with rich a priori knowledge may strike the ideal balance to make them effective and useful.

\section{Future Work \label{sec:futurework}}

Several of the scenarios that we considered in this paper were simplified versions of CTFs paired with basic RL algorithms; however, this setup has allowed us to make and illustrate our points clearly. Progress toward solving real-world problems would require, at the same time, scaling the complexity of CTFs and improving the way in which a RL agent manages structure and prior knowledge.

In terms of scaling structure, a direct way to achieve this would be to increase the sheer complexity of the problems by expanding the size of state and action space in order to resemble more closely what we see in reality. Complexity may also be increased by consistently adopting the assumption of \emph{non-stationarity}, as we briefly did in Simulation 2. 

In terms of learning the structure of the problem and integrate prior knowledge, better generalization (and scalability) can be achieved by switching from tabular algorithms to \emph{approximate algorithms}, thus sacrificing interpretability. 
More interestingly, it is possible to consider the possibility of learning through \emph{multiple channels} or relying on other forms of \emph{prior knowledge}; promising directions would be the integration of planning \cite{silver2017mastering}, hierarchical decomposition of a CTF in sub-tasks, reliance on \emph{relational inductive biases} \cite{battaglia2018relational}, or \emph{integration of logical knowledge} in the learning process \cite{besold2017neural}.

Tangentially, other challenges include the use of \emph{model learning} \cite{ha2018world}, in order to allow the agent to learn its own approximate model of the transition function of the environment, so that it could learn off-line via simulation; and proper \emph{reward shaping}, that is, providing rewards that may better guide the learning process. Finally, real-world agents may have to consider the problem of \emph{transfer learning} \cite{pan2009survey}, that is how to port the knowledge obtained from a class of CTF problems to another set of CTF problems.  

\section{Ethical considerations} \label{sec:Ethical}

Although this study considers and analyzes strengths and challenges of model-free RL agents on artificial problems, it is important to acknowledge that the development of real-world RL agents able to carry out actual PT presents the potential for malicious use. We would then like to stress that our results and suggestions are meant to foster the development of tools that may be of use to ethical hacker and find use in legitimate settings. We do not support and condemn the implementation of tools developed to attack and to harm, especially in a military context\footnote{\url{https://futureoflife.org/open-letter-autonomous-weapons/}}.

\section{Conclusions \label{sec:conclusions}}
In this work we considered CTF competitions as concrete instances of PT, and we modeled them as RL problems. We highlighted that a crucial challenge for a RL agent confronting a CTF problem is discovering a structure that is often limited and protected. We ran a varied set of simulations, implementing tabular Q-learning agents solving diverse CTF problems and exploring the contribution of injecting into the agent different forms of a priori knowledge. Our results confirmed the relevance of the challenges we identified, and we showed how different RL techniques (lazy loading, state aggregation, imitation learning) may be adopted to address these challenges and make RL feasible. 

We observed that while a strength of RL is its ability to solve model-free problems with minimal prior information, some forms of side information may be extremely useful for allowing the solution of a CTF in a reasonable time. We believe a constructive approach would be for RL to learn from standard artificial intelligence model-based methods and balance RL model-free inference with model-based deductions and inductive biases.

Our implementations are open and use standard interfaces adopted in the RL research community. It is our hope that this would make an exchange between the fields easier, with researchers in security able to borrow state-of-the-art RL agents to solve their problems, and RL researchers given the possibility of developing new insights by tackling the specific challenges instantiated in CTF games.

\ifCLASSOPTIONcaptionsoff
  \newpage
\fi



\bibliographystyle{IEEEtran}
\bibliography{IEEEabrv,ctf}
%



%


\begin{IEEEbiographynophoto}{Fabio Massimo Zennaro}
received his PhD in Machine Learning from the University of Manchester. Currently he is a postdoc researcher in the Digital Security and in the Oslo Analytics research groups at the University of Oslo. His research interests include representation learning, causality, fairness, and security.
\end{IEEEbiographynophoto}

\begin{IEEEbiographynophoto}{L{\'a}szl{\'o}~Erd{\H o}di}
has a PhD in cyber security. Currently he is a lecturer and researcher in the information security research group at the University of Oslo. His main research field is offensive security. He is the leader of the UiO Hacking Arena.  
\end{IEEEbiographynophoto}






\end{document}